\documentclass[conference]{IEEEtran}
\IEEEoverridecommandlockouts
\usepackage{cite}
\usepackage{amsmath,amssymb,amsfonts}
\usepackage[ruled,boxed,linesnumbered,vlined]{algorithm2e}
\usepackage{graphicx}
\usepackage{textcomp}
\usepackage{xcolor}

\usepackage{subcaption}

\usepackage{array}
\usepackage{makecell}

\usepackage{url}
\usepackage{pgfplots}
\pgfplotsset{compat=1.18}
\usepackage{pgfplotstable}
\usepackage{siunitx}
\usepackage{comment}
\usepackage{multirow}
\usepackage{setspace}
\usepackage{array}
\usepackage{wrapfig,booktabs}
\usepackage{xcolor, colortbl}
\newcommand{\rowcol}{\rowcolor{black!5}}
\usepackage{wrapfig}
\usepackage{graphicx}
\usepackage{subcaption}

\def\BibTeX{{\rm B\kern-.05em{\sc i\kern-.025em b}\kern-.08em
    T\kern-.1667em\lower.7ex\hbox{E}\kern-.125emX}}

\begin{document}
\title{MPBMC: \underline{M}ulti-\underline{P}roperty \underline{B}ounded \underline{M}odel \underline{C}hecking with GNN-guided Clustering}


\author{\IEEEauthorblockN{\small Soumik Guha Roy\IEEEauthorrefmark{1},  \small Sumana Ghosh\IEEEauthorrefmark{1}, \small Ansuman Banerjee\IEEEauthorrefmark{1}, \small Raj Kumar Gajavelly\IEEEauthorrefmark{3}, \small Sudhakar Surendran\IEEEauthorrefmark{4}}
\IEEEauthorblockA{\IEEEauthorrefmark{1} \footnotesize Indian Statistical Institute, Kolkata, India, Email: \{soumik\_r,sumana,ansuman\}@isical.ac.in, \\ 
\IEEEauthorrefmark{3} \footnotesize IBM Systems,India, Email: rgajavel@in.ibm.com, \IEEEauthorrefmark{4} \footnotesize Texas Instruments,India, Email: sudhakars@ti.com}
}

\maketitle

\begin{abstract}
Formal verification of designs with multiple properties has been a long-standing challenge for the 
verification research community. The task of coming up with an effective strategy that can efficiently cluster properties to be solved together has inspired a number of proposals, ranging from 
structural clustering based on the property cone of influence (COI) 
to leverage runtime design and verification statistics. In this paper, we present an attempt 
towards functional clustering of properties utilizing graph neural network (GNN) embeddings 
for creating effective property clusters. We 
propose a hybrid approach that can exploit neural functional representations of hardware circuits and 
runtime design statistics to speed up the performance of Bounded Model Checking (BMC) in the context of multi-property verification (MPV). Our method intelligently groups properties based on their functional embedding and design statistics, resulting in speedup in verification results. Experimental results on the HWMCC benchmarks show the efficacy of our proposal with respect to the state-of-the-art.
\end{abstract}

\begin{IEEEkeywords}
Multi-property Verification, Bounded Model Checking, Graph Neural Networks, Clustering
\footnote{This work is supported by Semiconductor Research Corporation (SRC).}
\end{IEEEkeywords}
\section{Introduction  \label{sec:intro}}

\begin{figure}[htbp]
\begin{tikzpicture}
\begin{axis}[
    width=9.9cm,height=3.9cm,
    xmin=0, xmax=100,
    xtick={0,20,40,60,80,100},
    ytick=\empty,
    xlabel={\# Frames},
    ylabel={\# CC},
    ylabel near ticks,
    xlabel near ticks,
    xlabel style={font=\small},  
    ylabel style={font=\small},  
    tick label style={font=\small},
    legend style={font=\scriptsize},
    legend pos=north west
]

\addplot[color=pink, line width=0.2pt] table[
    col sep=comma,
    x=x,
    y=y
] {Data/Motivational_Data/max_CC.csv};
\addlegendentry{Max CC}
\addplot[color=green, line width=0.2pt] table[
    col sep=comma,
    x=x,
    y=y
] {Data/Motivational_Data/avg_CC.csv};
\addlegendentry{Avg CC }

\addplot[color=blue, line width=0.2pt] table[
    col sep=comma,
    x=x,
    y=y
] {Data/Motivational_Data/cluster_CC.csv};
\addlegendentry{Random CC }

\addplot[color=red, line width=0.2pt] table[
    col sep=comma,
    x=x,
    y=y
] {Data/Motivational_Data/total_CC.csv};
\addlegendentry{Total CC }

\end{axis}
\end{tikzpicture}

\caption{Conflict Clause plot on 6s154.aig from HWMCC}
\label{fig:Random_Cluster_per_frame_conflict_clause}
\end{figure}
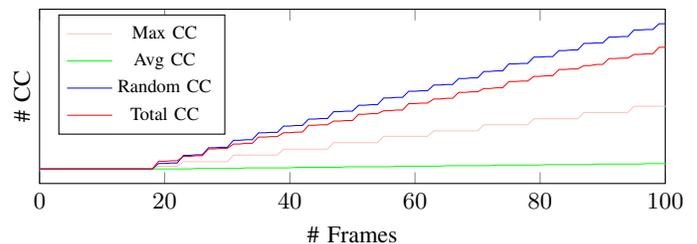

\noindent
Multi-property verification~\cite{cabodi_Model_Check_Multi_Prop} is a challenge for formal 
verification tools. 
On one extreme,  verification techniques that verify 
properties one at a time, lack information sharing and reuse, and thereby miss the chance to 
avoid redundant computations. On the other extreme, approaches that proceed with all properties 
together in tandem may suffer from a balancing concern, wherein hard-to-verify properties may slow 
down the overall progress. The key to solving this challenge is to create an effective property 
clusters consisting of properties that have high similarity and can be solved together. 
Each property depends on a specific part of the design’s logic, known as its \textit{cone of influence} (COI) or fanin logic; this includes all signals that 
directly or transitively affect the computation of the property status. 
\noindent
The insight behind solving  structurally / functionally similar properties 
together in a  verification run is to leverage the expected benefit that their proofs/counter-examples build along similar paths/substructures of the design, 
thereby avoiding redundant computations if the properties are taken up separately for verification. 
In this work, we consider Bounded Model Checking (BMC)~\cite{BMC_biere} verifiers that 
take in a design description along with a set of properties to be verified and 
incrementally unfold the design and the properties across increasing depths 
beginning from the start, in an attempt to falsify the constituent properties using 
a Satisfiability (SAT) solver to solve at every depth. 
An important aspect in SAT-based verification is 
Conflict Directed Clause Learning (CDCL)~\cite{CDCL_Book} wherein a SAT solver learns conflict clauses 
from one part of the design space, which can be effective in other parts of the 
search. In the context of BMC, as we proceed with multiple properties together, 
an important objective is to ensure that the learnt clauses benefit all properties 
in the given run. This necessitates the importance of an effective clustering strategy
which places functionally similar properties together and functionally different
ones in different runs. A random set of properties, if placed together in the 
same BMC run, may impede progress, since the properties may not be able to benefit 
from common conflict clauses (CC) and learnt clauses, 
thereby overwhelming the learnt clause repository, and 
slowing down the verification process. An example plot of CC versus increasing 
verification depth (also called frames in BMC parlance) is shown in 
Fig.~\ref{fig:Random_Cluster_per_frame_conflict_clause} 
on the design $6s154.aig$ selected from the Hardware Model Checking Competition (HWMCC)~\cite{hwmcc12,hwmcc13} benchmarks, where the number of conflict clauses (CC) using a random property cluster exceeds the individual CC count (both the average and maximum when properties are run separately). This increase in conflict clauses can make concurrent verification much slower than verifying properties one by one, thereby 
asserting the importance of an efficient clustering mechanism that can take in a 
given set of properties on a given design and create high affinity functional 
property groups, with a hope that 
when such properties are verified together, the overlap in their COIs can be exploited to enhance the verification via mutually beneficial clause learning, saving both time and computational resources. 

\noindent 
For effective multi-property verification, a number of research articles~\cite{Cabodi_Split_Group,Functional_Test_Generation_Mishra} in literature have proposed to create small property groups consisting of similar properties to be solved together, leveraging common subproblem sharing typically through structural grouping (COI-based) of properties, but with limited benefits. These approaches often overlook the semantic functional similarity between properties or runtime verification statistics. 
While clustering of properties based on functional similarity is helpful, it is computationally challenging to efficiently compute effective clusters that adequately capture functional similarities between properties. 
This work attempts to address these aspects towards clustering and effective verification of multi-property designs.

In this work, we use the graph neural network representation of circuits and 
property COIs to learn effective functionally similar property clusters.
The rise of deep learning (DL) has influenced several stages of the circuit design lifecycle~\cite{huang2021machine}, with representation learning emerging as a key technique for generating embeddings that support tasks like verification and high-level synthesis~\cite{DBLP:journals/todaes/FerrettiCZAP23}.
With the recent rise in successful deployments of Graph Neural Networks (GNN), there is a growing interest in the research community to use GNN models as circuit representations~\cite{dg2,dg} and use the same for EDA tasks like testability analysis~\cite{10.1145/3316781.3317838}, 
SAT solving~\cite{wang2024neuroback}. 
Despite the prevalence of GNN-based circuit representation methods, most of the circuit-based analysis approaches are dependent on the structural representation~\cite{cabodi2011optimized,dureja_boosting,dureja_accelerating}, runtime verification statistics~\cite{Purse}, synergies of BMC engines~\cite{memocode23,mab_bmc}. A recent work~\cite{BMC_Seq_VLSID_25,DeepIC3}, reports that the application of GNN-based circuit representations can speed up the verification.

The key contribution of our paper is to show that by learning functional representations of COIs of properties and leveraging runtime statistics jointly, it is possible to enhance the BMC-based circuit verification workflow by creating a judiciously chosen group of properties that are mutually beneficial and contribute to verification 
scalability. Our method has an offline data preparation phase, wherein we 
analyze and create overlapping property clusters based on neural embeddings of the 
COIs of 
properties on a database of available designs, that serves as close approximations of an exact functional similarity 
based clustering. At runtime for the design to be verified with multiple properties, we utilize runtime unfolding statistics of the given properties 
towards creating effective property groups. 
We consider the And-Inverter-Graph (AIG) representation of logic circuits and build our solution on top of the publicly available FV tool ABC~\cite{abc}, having support for AIG. We use the principle of inductive unfolding~\cite{inductive_unfolding}, which gives us better performance as reported in~\cite{BMC_Seq_VLSID_25}.
We conduct extensive experiments on the HWMCC benchmarks to show the performance of our GNN-aided multi-property design verification. We compare our performance against stand-alone property verification and a state-of-the-art technique based on clustering~\cite{dureja_boosting,dureja_accelerating}. We have the following contributions.
\begin{itemize}
    \item Use of GNN embeddings for effective property clustering.
    \item Use of property clusters in verification to speed up BMC.
    \item Extensive experimental results on HWMCC benchmarks.
\end{itemize}
\noindent
The rest of the paper is organised as follows. Section~\ref{sec:background}  provides a  background. 
Section \ref{sec:prob_statement} presents our methodology.
Experiments are in Section \ref{sec:result}. 
Section~\ref{sec:conc} concludes the paper.

\section{Background} \label{sec:background}
\noindent
\subsection{Multi-Property Bounded Model Checking} \label{sec:Formal_Modelling_Bounded_Model_Checking}

\noindent
In multi-property verification, 
given an unknown design $\widehat{\mathcal{U}}$, 
and a set of properties $S_P$, the model checker 
is tasked to verify if the design satisfies (or models) the properties. i.e.,  $ \widehat{\mathcal{U}} \models P, $ $\forall P \in S_P$.
A Bounded Model Checking (BMC) algorithm incrementally unfolds the design and the negation 
of a property to increasing depths (also called frames in BMC parlance) starting from the 
initial frame and searches for a counter-example (CEX) at each frame using a SATisfiability (SAT) solver. 
BMC tools are typically run with a 
maximum frame bound or time bound, within which it 
may either be able to conclude a property as {\em proven} (if the diameter of the design is reached), {\em falsified} (if it gets a counter-example at any depth) 
or report the status of a property as undetermined (UNDET), in which case it 
reports the maximum depth (also called the maximum unrolling depth) till which the property remained satisfiable and no counter-example could be found. 
The performance of a BMC engine is quantified either by the time taken to find the CEX (i.e., time metric) or by the maximum unrolling depth given a time bound if no CEX is found (i.e., unrolling depth metric). For a design with multiple properties, a BMC tool may work with one property at a time, 
or with all / subset of properties together in which case it proceeds breadthwise with all of 
the properties together. For individual runs, the time to CEX or maximum depth 
is reported, while for multiple properties together runs, the status of invididual properties (proven / falsified / UNDET) is reported along with the maximum unrolling depth. 

\subsection{Functional Representation Learning of Circuits}
\noindent 
Functional representation of a circuit 
focuses on the input output behavior of a circuit, 
thus showing better capabilities towards 
generalization of circuit behavior. 
Graph Neural Networks (GNN) offer a 
promising mechanism towards circuit representation learning. 
\noindent
{\em DeepGate}~\cite{dg,dg2} is a GNN model for obtaining a general and effective circuit representation. DeepGate embeds the logic function and structural information using attention mechanisms that mimics the logic procedure for circuit learning from unique circuit properties. DeepGate uses the ratio of logic-$1$ in the truth table as a functionality-related supervision metric. Due to inadequacy of the functionality-related supervision metric, multiple rounds of message passing are needed to preserve the logical correlation of gates, which is often time-consuming for large circuits. 
DeepGate2 (DG2)~\cite{dg2} attempts to counter these challenges,    
employing pairwise truth table differences using Hamming Distance between sampled logic gates as a supplementary supervision, with a loss function that minimizes the disparity between pairwise node embedding and pairwise truth table difference. 
Moreover, this method introduces an efficient one-round GNN that captures both structural and functional properties, leveraging the inherent circuit characteristics. We use DG2 model embeddings for circuit representation learning and property similarity evaluation. 


\section{Detailed workflow} \label{sec:prob_statement}
\noindent

\noindent
\noindent
Given an unknown design $\widehat{\mathcal{U}}$ with a set of $n$ properties, denoted as $S_P$, our objective 
is to create {\em high similarity property clusters} to be verified in the same BMC run. 
With these property clusters, we aim to maximize the BMC unrolling depth within a given time bound for each property for which a CEX is not found, or minimize the time taken to generate a CEX otherwise. 
To build high similarity property clusters, we use popular machine learning algorithms for 
clustering on the functional tensors of the COIs of the 
properties. We assume we have a database  ${\mathcal{C}}$ 
of already verified designs, each with multiple properties, for which the best performing clustering information is available when we verify $\widehat{\mathcal{U}}$. We use this information for
creation of property clusters for $\widehat{\mathcal{U}}$.
\noindent

\begin{figure}[!t]
    \centering
    \includegraphics[width=\linewidth,height=2.0in]{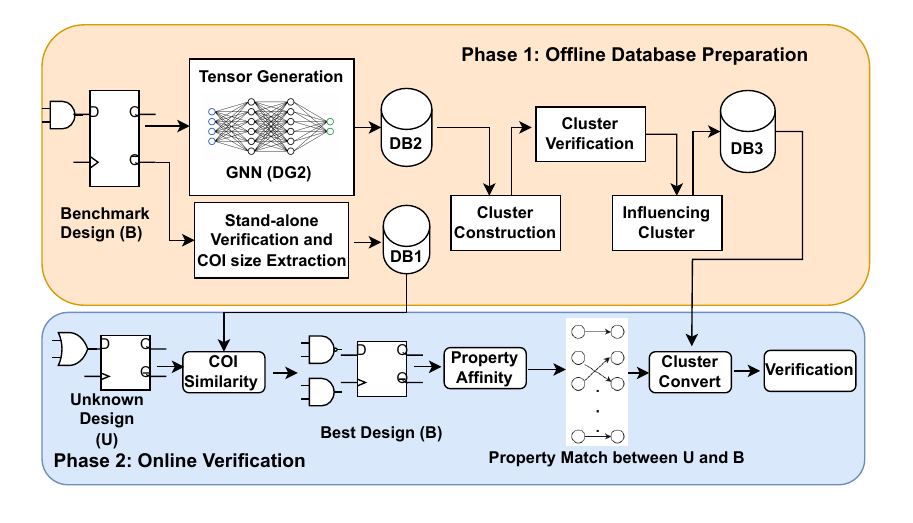}
    \caption{Our Framework - MPBMC}
\label{fig:MPEG_Framework}
\end{figure}

Our framework, shown in Fig.~\ref{fig:MPEG_Framework} consists of two main phases, (a) an offline database preparation phase on ${\mathcal{C}}$ and (b) an online phase with the objective of efficient verification for $\widehat{\mathcal{U}}$ guided by a 
design $\mathcal{B} \in {\mathcal{C}}$ from offline information. The overall approach for the proposed framework is as below. 
In the offline phase, verification status and COI sizes of each property are collected for each reference circuit in $\mathcal{C}$. GNN-based embeddings identify functionally similar properties, which are clustered and selectively retained based on verification results. In the online phase, the unknown design $\widehat{\mathcal{U}}$ is matched to a similar benchmark $\mathcal{B} \in \mathcal{C}$ using COI-based similarity, and the top-performing clusters from $\mathcal{B}$ guide the verification of $\widehat{\mathcal{U}}$. The details of the proposed framework follow next.

\subsection{Offline Data Preparation}
\label{subsec:offline_Data_Preparation}
\noindent
This offline phase is done once to prepare three databases, $\mathcal{DB}_1$, $\mathcal{DB}_2$, and $\mathcal{DB}_3$ based on the set of given benchmark circuits $\mathcal{C}$.
$\mathcal{DB}_1$ stores the circuit's structural data, $\mathcal{DB}_2$ holds property embeddings of each design, and $\mathcal{DB}_3$ stores influencing cluster information 
to guide the search for the best property group to be used for faster verification for design $\mathcal{\widehat{U}}$. We explain the construction of each below. 
\noindent
For each design in $\mathcal{C}$, we carry out the following steps to prepare the databases.

    {\em Single Property Run and Information Collection: } 
    We perform a standalone BMC of each property of the design for a fixed $\mathcal{T}$ amount of time and store the verification metric value (time / unrolling depth), verification status (SAT / UNSAT / UNDET), and property COI size information. 
    We store this information in $\mathcal{DB}_1$ for use at a later stage.

    {\em Tensor Generation: } In this phase, we find the functional representation of each property using the GNN embedding of each property COI. 
    This is utilized in the cluster construction phase, with the representations 
    helping us move to a different embedding space to perform cluster-based similarity analysis between properties. 
    For creating these GNN embeddings, we use DG2. It may be noted that DG2 works only on combinational designs; thus, for a sequential circuit $\mathcal{B} \in \mathcal{C}$, we generate an inductive unfolding to get a combinational circuit netlist and generate the tensor. These GNN embeddings constitute the database $\mathcal{DB}_2$. 

    {\em Cluster Construction: } Once the GNN embeddings of the COI of 
    each property are available, we use K-means and K-medoids clustering
    to identify high-affinity groups of properties, based on their functional embeddings generated in the previous step. It may be noted that the cluster sets formed $S_{cluster}$=$\{S_1,S_2,\dots,S_n\}$ where each $S_i$ is a set of 
    {\em functionally similar} properties, with sizes ranging from $2$ to the total number of properties in the design, are not necessarily mutually disjoint. 
    This is what makes our approach distinctly different from other approaches 
    that cluster properties into disjoint clusters. The unique idea that drives 
    our approach is that a property may belong to multiple clusters; it is 
    solved in one cluster where it gains the most, and its presence helps to 
    expedite the solution of others in the other clusters where it is present. 
    We explain this in detail at a later stage.

    {\em Cluster Execution and Information Collection: } This phase has a significant effect on the overall performance as it guides the detection of influencing clusters (defined next) along with their influenced properties. With the cluster configurations obtained above, we carry out a BMC run on {\em each 
    cluster} separately with the design, and only the set of properties that 
    belong to the cluster. The time allotted to this BMC run is $\mathcal{T} 
    \times$ {\em cluster size}, where cluster size is in terms of the number of 
    properties present in the 
    cluster, and $\mathcal {T}$ is the time value used above for a single property run.

    {\em Influencing Cluster Identification: } For each property, we now identify which cluster is the best performing one. 
    As mentioned earlier, our clusters are not disjoint; thus, a property may belong to multiple clusters. We aim to identify which among these clusters gives the best benefit in performance in terms of a gain metric, which we define next. 
    We refer to this as the {\em influencing cluster} for the property. 
    The influencing cluster of a property depends on the verification result of the single property run (done above) and the runs of the clusters in which it belongs. The verification result of a property might change while verifying the same using a cluster of properties. Based on the transition of verification status and the property evaluation metric described below, the influencing cluster identification is done. 
We define the concept of {\em property gain} 
which is the basis behind this step.
To identify the influencing cluster for a property, we need to quantify the gain of each property $P_i$, i.e., $Gain(P_i)$, when it is verified alone versus when it is verified in a cluster. With the results of the stand-alone and cluster execution obtained above, we have 5 possible transitions in the verification status, as below.
\begin{itemize}

    \item UNDET to SAT: This indicates a property $P_i$ is UNDET in a stand-alone run and becomes SAT in a cluster run with $n$ other properties. If it takes $t_c$ time to derive SAT, then the gain for $P_i$ is taken as $Gain(P_{i})$ = $t_c/n$.
    
    \item SAT to SAT: Here, $P_i$ is SAT in both stand-alone and cluster run, assume the corresponding times to derive SAT are $t_s$ and $t_c$ respectively. In this case, the gain for $P_i$ is taken as $Gain(P_i)$ = $(t_s - t_c)/t_s$. 
    Note that this is the relative change in SAT time with respect to its single run. 
    A negative gain value indicates $t_s < t_c$.
    
    \item UNDET to UNDET: In this case, the verification status is UNDET in both the standalone and cluster runs above. Let $d_s$ and $d_c$ denote the maximum unrolling depth 
    in a given time bound in standalone and grouped verification for $P_i$. 
    Now the gain value for $P_i$ is $Gain(P_i)$ = $(d_c - d_s) / d_s$. A positive gain indicates that $d_c > d_s$. 
    
    \item SAT to UNDET: This indicates that a property is SAT in a stand-alone run but UNDET in cluster verification, which means that the presence of a set of properties degrades the verification of $P_i$. The gain is defined like the UNDET-to-UNDET case, as $Gain(P_i)$ = $(d_c - d_s)/d_s$. 
    
    \item UNSAT to UNDET: Here, the property is found to be UNSAT in stand-alone verification but UNDET while verifying in a cluster. 
    The gain $Gain(P_i)$ is defined in the same way as in the SAT to UNDET case.

    \item UNDET to UNSAT: Similar to the UNDET to SAT case, this transition indicates that a property $P_i$, which is UNDET in a stand-alone run, becomes UNSAT when verified in a cluster with other properties. The gain for $P_i$ is defined analogously to the UNDET to SAT case.
    
    
\end{itemize}
\noindent
UNDET-to-SAT and UNDET-to-UNSAT cases are most desirable, while UNSAT-to-UNDET is not desirable. With the information above, we identify the influencing cluster for each property.
As discussed in the cluster construction phase, a property $P_i$ $\in$ $S_j$ $\in$ $S_{cluster}$ 
can belong to multiple clusters, with different cluster-specific gains, denoted as $Gain(P_i)_{S_j}$. 
Among the associated clusters of $P_i$, the cluster $S_I$ having the highest gain for $P_i$ is the {\em influencing cluster} or top performing cluster for $P_i$. 
\noindent
$P_i$ is termed as influenced by $S_I$.
To identify the influencing cluster $S_I$ 
for a property $P_i$, we need to compare the gain values for $P_i$, considering all clusters to which it 
belongs. The gain value for a property with respect to a cluster is a 6-element vector, with a non-zero entry in one of the positions only. For this comparison, we need to set up an order between the 6 fields above. 
We use the total order: UNDET-to-SAT $=$ UNDET-to-UNSAT $>$ SAT-to-SAT $>$ UNDET-to-UNDET $>$ SAT-to-UNDET $>$ UNSAT-to-UNDET. 
Consider an example property $P$ which belongs to $4$ different clusters $\{S_1,S_2,S_3,S_4\}$, resulting in $4$ different cluster-specific gain values as $Gain(P)_{S_i}$ where $i \in \{1,2,3,4\}$. 
For example, a cluster $S_2$ is an influencing cluster for the property $P$ if $Gain(P)_{S_2} > Gain(P)_{S_i} $, where $i \in \{1,3,4\}$ considering a vector to vector comparison. It may happen that due to different transitions in verification status, cluster-specific gains of $P$ are different. In that case, the metric evaluation order, discussed above, is used to find the influencing cluster for $P$.
To illustrate the evaluation order with an example, assume a property $Q$ is associated with two different clusters $S_1$ and $S_2$. The transition of verification status of $Q$ in $S_1$ and $S_2$ is UNDET-to-UNDET and UNDET-to-SAT, respectively. Let the cluster specific gain for the property $Q$ are $Gain(Q)_{S_1}=0.9$ and $Gain(Q)_{S_2}=0.4$. Although the cluster-specific gain for $S_2$ is less, but metric evaluation order gives priority to $S_2$ to be the influencing cluster for $Q$.

The setup of $\mathcal{DB}_3$ is completed hereafter. As an output of the above, we have identified the influencing cluster for each property in a design. $\mathcal{DB}_3$ contains this information 
for all designs, for all constituent properties.
As an example, consider a design $\mathcal{B}$ with the following 
clusters $\{\{P_1,P_4\},\{P_1,P_2,P_3\},\{P_2,P_3,P_4\},\{P_2,P_3\},\{P_1,P_3,P_4\}\}$. Table~\ref{Influencing_Cluster_Design_Table} shows the influencing clusters for $\mathcal{B}$.

\begin{table}[!h]
\footnotesize
\begin{center}
\caption{Influencing clusters for a design $\mathcal{B}  $} \label{Influencing_Cluster_Design_Table}
\begin{tabular}{|>{\centering\arraybackslash}p{1.5cm}|>{\centering\arraybackslash}p{1.4cm}|>{\centering\arraybackslash}p{0.9cm}|>{\centering\arraybackslash}p{1.5cm}|>{\centering\arraybackslash}p{0.9cm}|} 
 \hline
 \textbf{Property} & $P_1$ & $P_2$ & $P_3$ & $P_4$ \\ 
 \hline
 \textbf{Influencing Cluster} & $\{P_1,P_3,P_4\}$ & $\{P_2,P_3\}$ & $\{P_1,P_3,P_4\}$ & $\{P_1,P_4\}$ \\ 
 \hline
\end{tabular} 
\end{center}
\end{table}
\noindent

\subsection{Online verification} 
\label{method}

\noindent
\noindent
Given an unknown circuit $\mathcal{\widehat{U}}$ that has $\mathcal{P}_{\widehat{U}}$ number of properties to be verified using BMC, we consider the time bound as $\mathcal{T} \times \mathcal{P}_{\widehat{U}}$, where $\mathcal{T}$ is the fixed verification time bound (Sec.~\ref{sec:prob_statement}) given to each property in $\mathcal{\widehat{U}}$. We now use the databases $\mathcal{DB}_1$ and $\mathcal{DB}_3$ to use the influencing cluster to accelerate the verification of $\widehat{\mathcal{U}}$.
Our goal is to select the most similar design $\mathcal{B}$ from the database of known designs $\mathcal{C}$, whose influencing cluster information will guide the BMC for verification of $\widehat{\mathcal{U}}$. 
\begin{algorithm}[hbt!]
\small
\caption{\label{algo:online_algo}Online Verification}
\KwIn{Unknown design $\widehat{\mathcal{U}}$, Databases $\mathcal{DB}_1$, $\mathcal{DB}_3$,threshold $\delta$}

$DB_{\text{pruned}} \gets \Delta_{\text{design\_info}} \gets \emptyset$ \; 

\ForEach{design $\in \mathcal{DB}_1$}{
    \If{$P_{\widehat{\mathcal{U}}} - \delta < P_{\text{design}} < P_{\widehat{\mathcal{U}}} + \delta$}{
        $\mathcal{DB}_{\text{pruned}} \gets \mathcal{DB}_{\text{pruned}} \cup \{\text{design}\}$ \;
    }
}




\ForEach{ design $\in \mathcal{DB}_{\text{pruned}}$}{
    
    $design\_info_1 \gets \texttt{getInfo}(design, \mathcal{DB}_1)$ \tcp*{Returns COI Size Info}
    $design\_info_2 \gets \texttt{getInfo}(\widehat{\mathcal{U}}, \mathcal{DB}_1)$ \tcp* {Returns COI Size Info After Unfolding}

    $diff \gets \left| design\_info_1 - design\_info_2 \right|$    \tcp* {Finding COI Difference}
    
    $\Delta_{\text{design\_info}} \gets \Delta_{\text{design\_info}} \cup \texttt{}{(design, diff)}$ \;
}

$\mathcal{B}  \gets \min (\Delta_{\text{design\_info}})$ \tcp*{Most Similar Design}

\ForEach{$prop_1 \in \mathcal{B}$}{
    \ForAll{$prop_2 \in \widehat{\mathcal{U}}$}{
        $info_{P_1} \gets \texttt{getInfo}(\mathcal{P}_1, \mathcal{D}_{\text{depth}}, \mathcal{DB}_1)$ \tcp*{Find property COI after unfolding}
        $info_{P_2} \gets \texttt{getInfo}(\mathcal{P}_2, \mathcal{D}_{\text{depth}}, \mathcal{DB}_1)$ \;

        $diff\_Matrix[i][j] \gets \left| info_{P_1} - info_{P_2} \right|$ \tcp* {Store Property COI diff in Matrix}
    }
}
$UD_{\text{Map}} \gets \texttt{property\_association}(diff\_Matrix)$ \tcp* {Find Property Association in terms of Smaller COI Diff}


$inf_{\text{cluster}} \gets \texttt{getInfCluster}(\mathcal{B}, \mathcal{DB}_3)$ \tcp* {Find Influencing Cluster}
\tcc{Cluster Conversion Starts}
$\mathcal{C'}=\emptyset$ \;
\ForEach{$cls \in inf_{\text{cluster}}$}{
    
    $cls'=\phi$ \;
    \ForEach{$p \in cls$} {
        $p'=UD_{\text{Map}}(p)$ \;
        $cls'=cls' \cup{ p'}$ \;       
    }
    
    $\mathcal{C'}=\mathcal{C'} \cup{ cls'}$ \;
}

\tcc{Verification Starts}
\texttt{Verify}($\mathcal{C'}$) \;
\end{algorithm} 
This consists of two main steps. 

{\em Initial Pruning:} From $\mathcal{DB}_1$, we first 
extract the set of circuits $\mathcal {S}$ whose property count falls in the range 
$[ \mathcal{P_{\widehat{U}}}-\delta, \mathcal{P_{\widehat{U}}}+\delta]$, where $\delta > 0$ is a tuning parameter that reduces the overall search space to a limited number of known circuits having a total number of properties close to $\mathcal{P_{\widehat{U}}}$.

{\em Similarity Search and Cluster Construction:} This part consists of three main steps as described below.
\begin{itemize} 
        \item {\textit{Most Similar Circuit Identification}: } First, we compare the design information (number of gates, sequential elements, 
        inverters) of all the circuits in $\mathcal{S}$ with $\widehat{\mathcal{U}}$ 
        using the database 
        $\mathcal{DB}_1$. The circuit having minimal difference in COI size information 
        is used as the most similar circuit $\mathcal{B}$.
        
        \item {\textit{Property Mapping}: } Next, we find the similarities between 
        properties in $\mathcal{B}$ and $\widehat{\mathcal{U}}$. For this, we unfold each property of $\widehat{\mathcal{U}}$, extract its COI size, and compare it with the COI size of each property of $\mathcal{B}$, stored in $\mathcal{DB}_1$. We find the differences in the information about the COI size between each pair of properties, and this gives us, for each property 
        in $\widehat{\mathcal{U}}$, the most similar property in $\mathcal{B}$. 
        
        \item {\textit{Influencing Cluster Information Extraction}: } Next, the influencing cluster information for each property in $\mathcal{B}$ is extracted from $\mathcal{DB}_3$. Now, the extracted influencing clusters are converted to the corresponding cluster of properties using the 
        properties of $\widehat{\mathcal{U}}$.
        
\end {itemize}
\noindent
Once the clusters are obtained for $\widehat{\mathcal{U}}$, BMC is run 
one cluster at a time, with all the properties in each cluster running together, with 
a total time per cluster as $\mathcal{T} \times$ {\em cluster-size} as explained earlier. 
We record the verification status and time / unrolling depth for each cluster run, and compare these values with stand-alone runs of each property and a recent work, as 
explained in the following. 
Algorithm~\ref{algo:online_algo} summarizes the overall approach.

\section{Experimental Evaluation}
\label{sec:result}
\begin{table*}[!h]
\centering
\caption{Comparative results on HWMCC`12-13 Benchmark}
\label{tab:table_UNDET_multi_prop}
\footnotesize
\renewcommand{\arraystretch}{1}
\begin{tabular}{|l|c|c|c|c|c|ccc|ccc|}
\toprule
\multirow{2}{*}{\bf Benchmark} & 
\multirow{2}{*}{\bf \#AND} & 
\multirow{2}{*}{\bf \#LAT} & 
\multirow{2}{*}{\bf \#Prop} & 
\multirow{2}{*}{\bf \#Inf-Prop} & 
\multirow{1}{*}{\bf \#Prop } &  
\multicolumn{3}{c|}{\bf MPBMC UNDET Gain (\%)} &
\multicolumn{3}{c|}{\bf ~\cite{dureja_boosting,dureja_accelerating} UNDET Gain (\%)} \\
\cline{7-12}
 & & & & & {\bf Gainer(\%)} & {\bf Min} & {\bf Avg} & {\bf Max} & {\bf Min} & {\bf Avg} & {\bf Max} \\
\midrule
\ttfamily{6s154} & 18183 & 128 & 32 & 29 & 76 & 1.81 & 238.17 & 1200 & 285.71 & 1110.98 & 2600\\
\rowcolor[gray]{0.9}\ttfamily{6s329} & 1691790 & 31947 & 33 & 29 & 93 & 2.38 & \textcolor{red}{13.61} & \textcolor{red}{34.37} & \textcolor{teal}{-} & \textcolor{teal}{-} & \textcolor{teal}{-} \\
\ttfamily{6s343} & 43996 & 4774 & 49 & 49 & 59 & 1.29 & 43.69 & 138.98 & \textcolor{teal}{-88.14} & \textcolor{teal}{-44} & \textcolor{teal}{36.36}\\
\rowcolor[gray]{0.9}\ttfamily{bob12m07m} & 63503 & 1258 & 53 & 50 & 68 & 4 & 20.34 & 50 & \textcolor{teal}{-11.76} & \textcolor{teal}{0.36} & \textcolor{teal}{12.5}\\
\ttfamily{6s340} & 42291 & 3337 & 76 & 57 & 51 & 3.22 & 20.22 & 64.75 & \textcolor{teal}{-2.33} & 26.61 & \textcolor{teal}{55.56}\\
\rowcolor[gray]{0.9}\ttfamily{6s305} & 36517 & 8000 & 116 & 43 & \textcolor{red}{36} & 2 & 45.45 & 183.82 & \textcolor{teal}{-88.49} & 234.11 & 5065.16 \\
\ttfamily{nusmvdme1d16multi} & 1616 & 321 & 120 & 114 & 85 & \textcolor{red}{0.76} & 18.08 & 46.36 & \textcolor{teal}{-16.78} & \textcolor{teal}{5.66} & \textcolor{teal}{37.38}\\
\rowcolor[gray]{0.9}\ttfamily{bob12m01m} & 74234 & 6555 & 142 & 98 & 80 & 5.89 & 53.5 & 164.28 & \textcolor{teal}{-1.43} & \textcolor{teal}{-13.33} & \textcolor{teal}{7.69}\\
\ttfamily{6s421} & 6294 & 951 & 150 & 89 & \textcolor{blue}{100} & \textcolor{blue}{21.90} & \textcolor{blue}{430.23} & \textcolor{blue}{2647.76} & \textcolor{teal}{21.91}& \textcolor{teal}{404.44} & \textcolor{teal}{2367.16}\\
\rowcolor[gray]{0.9}\ttfamily{6s409} & 120917 & 10523 & 184 & 122 & \textcolor{blue}{100} & 22 & 100.75 & 354.55 & 69.6 & 158.74 & \textcolor{teal}{345.07}\\
\bottomrule
\end{tabular}
\fcolorbox{black}{blue} {\rule{0pt}{4pt}\rule{3pt}{0pt}} \text{Max gain}
\fcolorbox{black}{red} {\rule{0pt}{4pt}\rule{3pt}{0pt}} \text{Min Gain}
\fcolorbox{black}{teal} {\rule{0pt}{4pt}\rule{3pt}{0pt}} \text{Same or higher gain than SOTA}  
\end{table*}

\noindent
{\em Benchmarks and Setup: }
To show the efficacy of the proposed framework, we conducted experiments on a set of $30$ benchmark circuit designs taken from the HWMCC 2012-13~\cite{hwmcc12,hwmcc13} 
benchmarks. The number of AND gates and latch count of these designs lie within the range of $(1.5k - 16k)$ and $(65 - 102k)$, respectively. Out of these $30$ designs, $20$ designs were used to prepare our $3$ databases $\mathcal{DB}_1$, $\mathcal{DB}_2$ and $\mathcal{DB}_3$. We applied the Principal Component Analysis (PCA) during $\mathcal{DB}_2$ preparation with $95 \%$ of significant variations preserving the design embeddings. The remaining $10$ designs were considered as the new (unknown) circuits $\widehat{\mathcal{U}}$ on which the verification was done. Note that for all these $20$ designs, we set the timeout time $\mathcal{T}$ to $15$ minutes as discussed in Section~\ref{sec:prob_statement}. 
The details (Circuit Name, AND gate, Latch counts, number of properties) of these $10$ designs are reported in Columns 1-3 of Table ~\ref{tab:table_UNDET_multi_prop}. 
In this work, \texttt{bmc3g} is considered as the underlying BMC engine which is available as part of the verification tool, ABC~\cite{abc}. All experiments were carried out on an Intel 14700K processor, 16GB RAM, NVIDIA T100 8GB GPU. For GNN embeddings, we used DG2~\cite{dg2} version 2.0.1. The total size of $\mathcal{DB}_1,\mathcal{DB}_2$ and $\mathcal{DB}_3$ is nearly $600$ MB.  Additionally, the offline data preparation phase required an average of $210$ seconds per design for functional embedding generation. The subsequent clustering step took an average of $28$ seconds and $121$ seconds per design using the k-means and the k-medoids algorithm, 
respectively.

{\em Result Analysis: }
Since all the benchmark designs considered are of UNDET type (as reported in HWMCC reports), we consider UNDET (i.e., maximum unrolling depth) as the metric for the performance of BMC verification. Table ~\ref{tab:table_UNDET_multi_prop} reports the respective results where the maximum gain across properties ranges between $34.37 \%$ to $2647.76\%$ while the average gain ranges between $13.61 \%$ to $430.23 \%$. We consider only those properties as influenced properties, where the UNDET depth in stand-alone verification is less than its associated cluster verification. In summary, we see that more than $51\%$ of total influenced properties denoted as $\textit{Inf-Prop}$ in Table~\ref{tab:table_UNDET_multi_prop} (Col-V) achieve higher UNDET depth gain shown as $Prop$ $Gainer$ in Col-VI. We show that our gain achieved is better than the state-of-the-art method~\cite{dureja_boosting,dureja_accelerating} (Col X-XII), where we take the reported clusters and compare them with our framework. 
We achieve higher UNDET depths in $8$ among the $10$ cases. Notably, for the design $6s329$ having maximum AND gate and latch count, our method successfully forms property clusters and outperforms~\cite{dureja_boosting,dureja_accelerating}, which fail to generate any cluster.
\begin{figure}[htbp]
\begin{tikzpicture}
\begin{axis}[
    width=9.9cm,height=3.9cm,
    xmin=0, xmax=100,
    xtick={0,20,40,60,80,100},
    ytick=\empty,
    xlabel={\# Frames},
    ylabel={\# CC},
    ylabel near ticks,
    xlabel near ticks,
    xlabel style={font=\small},  
    ylabel style={font=\small},  
    tick label style={font=\small},
    legend style={font=\scriptsize},
    legend pos=north west
]

\addplot[color=pink, line width=0.2pt] table[
    col sep=comma,
    x=x,
    y=y
] {Data/UD_6s409/max_CC.csv};
\addlegendentry{Max CC}

\addplot[color=green, line width=0.2pt] table[
    col sep=comma,
    x=x,
    y=y
] {Data/UD_6s409/min_CC.csv};
\addlegendentry{Min CC}
\addplot[color=red, line width=0.2pt] table[
    col sep=comma,
    x=x,
    y=y
] {Data/UD_6s409/all_prop_singlerun_max_CC.csv};
\addlegendentry{Total CC }
\addplot[color=blue, line width=0.2pt] table[
    col sep=comma,
    x=x,
    y=y
] {Data/UD_6s409/cluster_CC.csv};
\addlegendentry{MPBMC CC }

\end{axis}
\end{tikzpicture}

\caption{Conflict Clause (CC) Analysis}
\label{fig:per_frame_conflict_clause}
\end{figure}
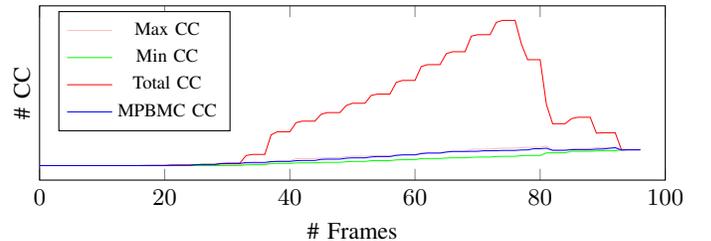

\noindent
{\em Conflict Clause Analysis:} \label{sec:per_frame_conflict_clause_analysis}
Figure~\ref{fig:per_frame_conflict_clause} shows that for the design $6s409$, conflict clause generation per frame in MPBMC is significantly less than in stand-alone verification. This result justifies that mutually beneficial clause learning among the grouped properties helps in reducing conflict clause generation, thereby reducing verification effort.

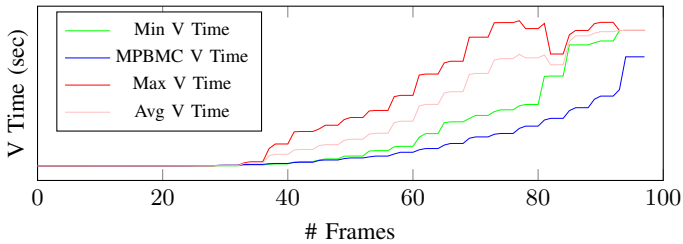
\begin{figure}[htbp]
\begin{tikzpicture}
\begin{axis}[
    width=9.9cm,height=3.9cm,
    xmin=0, xmax=100,
    xtick={0,20,40,60,80,100},
    ytick=\empty,
    xlabel={\# Frames},
    ylabel={V Time (sec)},
    ylabel near ticks,
    xlabel near ticks,   
    xlabel style={font=\small},  
    ylabel style={font=\small},  
    tick label style={font=\small},
    legend style={font=\scriptsize},
    legend pos=north west
]

\addplot[color=green, line width=0.2pt] table[
    col sep=comma,
    x=x,
    y=y
] {Data/UD_6s409/minTime.csv};
\addlegendentry{Min V Time}

\addplot[color=blue, line width=0.2pt] table[
    col sep=comma,
    x=x,
    y=y
] {Data/UD_6s409/clusterTime.csv};
\addlegendentry{MPBMC V Time}
\addplot[color=red, line width=0.2pt] table[
    col sep=comma,
    x=x,
    y=y
] {Data/UD_6s409/maxTime.csv};
\addlegendentry{Max V Time}
\addplot[color=pink, line width=0.2pt] table[
    col sep=comma,
    x=x,
    y=y
] {Data/UD_6s409/avgTime.csv};
\addlegendentry{Avg V Time}
\end{axis}
\end{tikzpicture}
\caption{Verification Time (V Time) Comparison}
\label{fig:per_frame_verific_time}
\end{figure}

{\em Verification Time Analysis:} \label{sec:per_frame_verific_time_analysis}
Here, we consider a cluster of the same design with $11$ properties where $10$ properties have an average $141.85\%$ UNDET gain.  Figure~\ref{fig:per_frame_verific_time} indicates that the verification time of our proposed MPBMC method for the influencing cluster is less than all other verification times of all the constituent properties using stand-alone verification.

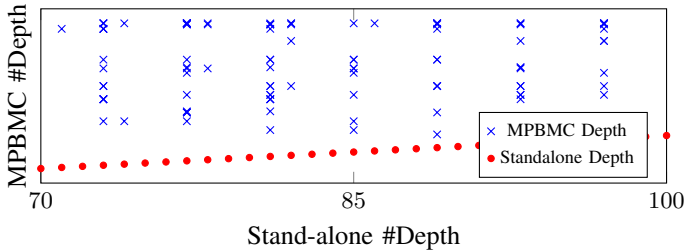
\begin{figure}[htbp]
\begin{tikzpicture}
\begin{axis}[
    width=9.9cm,height=3.9cm,
    xmin=70, xmax=100,
    xtick={70,85,100},
    ytick=\empty,
    xlabel={Stand-alone \#Depth},
    ylabel={MPBMC \#Depth},
    ylabel near ticks,
    xlabel near ticks,
    tick label style={font=\small},
    legend style={font=\scriptsize},
    legend pos=south east
]

\addplot[color=blue,only marks,mark size=2pt,mark=x] table[
    col sep=comma,
    x=x,
    y=y
] {Data/UD_6s409/single_cluster.csv};
\addlegendentry{MPBMC Depth}

\addplot[color=red, only marks,mark size=1.2pt] table[
    col sep=comma,
    x=x,
    y=y
] {Data/UD_6s409/base_depth.csv};
\addlegendentry{Standalone Depth}
\end{axis}
\end{tikzpicture}
\caption{Stand-alone vs MPBMC UNDET Depth Comparison}
\label{fig:UNDET_depth_analysis}
\end{figure}

{\em UNDET Depth Analysis:} \label{sec:undet_depth_analysis}
Fig.~\ref{fig:UNDET_depth_analysis} shows that all properties of the same design $6s409$ achieve higher UNDET depth using our proposed method compared to the standalone one.

\section{Conclusion and Future Work \label{sec:conc}}
\noindent
This work proposes a method for multiproperty verification using 
clustering and offline design information stored as tensor embeddings. 
Verifying functionally similar properties together helps in developing mutually beneficial conflict clauses and thus expedites verification. 
Going ahead, we plan to explore advanced clustering methods that leverage subproblem sharing between properties and incorporate dynamic clustering, where cluster configurations adapt over time based on runtime verification statistics. This enhances the effectiveness further and strengthens the overall verification flow. Additionally, we intend to perform a detailed sensitivity analysis of key hyperparameters such as cluster size, GNN architecture, and similarity thresholds to better understand their impact on verification efficiency and accuracy. 


\bibliographystyle{ieeetr}
	\bibliography{9_reference}
\end{document}